\documentclass[twocolumn]{aastex631}

\begin{document}

\title{Understanding planet formation using
microgravity experiments \footnote{Published in Nature Reviews Physics, 3, 405-421, 2021. https://www.nature.com/articles/s42254-021-00312-7}}

\author[0000-0002-7962-4961]{Gerhard Wurm}
\affiliation{University of Duisburg- Essen\\
Faculty of Physics\\
Duisburg, Germany}

\author{Jens Teiser}
\affiliation{University of Duisburg- Essen\\
Faculty of Physics\\
Duisburg, Germany}



\begin{abstract}
\noindent
In 2018, images were released of a planet being formed around the star PDS 70,
offering a tantalizing glimpse into how planets come into being. However, many questions remain
about how dust evolves into planets, and astrophysical observations are unable to provide all
the answers. It is therefore necessary to perform experiments to reveal key details and, to avoid
unwanted effects from the Earth’s gravitational pull, it is often necessary to perform such
experiments in microgravity platforms. This Review sketches current models of planet formation
and describes the experiments needed to test the models.

\end{abstract}



\section{Introduction} \label{sec:intro}
\noindent
Planet formation is not an abstract, far-off astrophysical
concept: we live on a well-situated terrestrial planet.
As such, it is interesting to try to explain how planetary
systems come into existence. At the star PDS 70, a
planet still embedded in its cradle - a protoplanetary
disk - has been detected \citep{Keppler2018,Haffert2019} (Fig. \ref{fig:1}), giving the tantalizing
impression that we might actually see how planets,
including Earth, come to life. But although we directly
see the dust in these disks as initial conditions, and we
image exoplanets as final stages of planet formation,
we are blind to the intermediate phases of how the dust
evolves into planets. Understanding these intermediate
phases requires theory, numerical simulations and
laboratory experiments. The latter often act as judges
on proposed mechanisms of pre-planetary evolution.
Many such dedicated ‘planet formation experiments’
have been carried out. Those under the Earth’s gravity
cannot answer all our questions, and therefore
microgravity is often mandatory.
This Review outlines the basic steps in current models
of planet formation. It starts from dust and ends at
planets. We put emphasis on the problematic phases and
on the conditions under which planets and their precursors
form. More specifically, we consider the interaction
of particles, with its many aspects of collisions,
but also interactions with the surrounding gas and radiation.
We pick up specific phases in planet formation,
the experiments that have been carried out to simulate
these phases - ground-based but especially microgravity
studies - and what has been learned from them, and
where this research might lead to in the future.

\section{Witnesses to planet formation} \label{sec:witness}
\begin{figure}
\epsscale{1.2}
\plotone{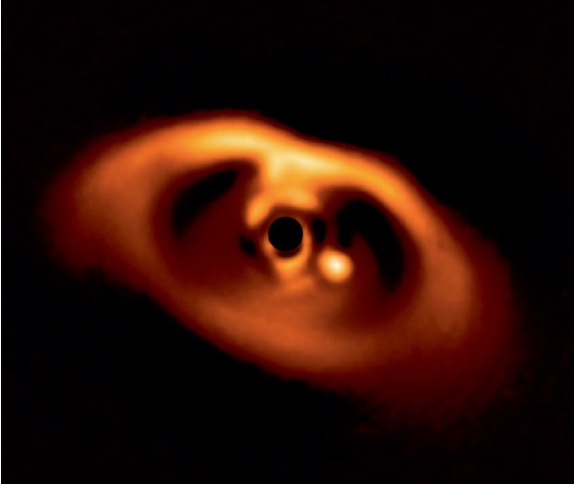}
\caption{\small{{PDS 70, a young star $\sim$5 million years old with a planetary system in the
making, where a planet embedded in a protoplanetary disk is directly imaged.}
This image was taken by the SPHERE instrument at the Very Large Telescope. The planet
is the bright spot on the right from the centre about 20 astronomical units from the star
(masked by the black circle) \citep{Keppler2018,Mueller2018}. Credit: ESO/A. Müller et al.}}
\label{fig:1}
\end{figure}

\noindent
Being rather close, the bodies of the Solar System are the
witnesses to planetary formation that we can study best.
Observing them offers some initial insights into how
they formed. Small bodies such as asteroids and comets
are promising sources of information, as they never
made it to full-size planets. In 2014–2016, the Rosetta
mission visited comet 67P/Churyumov–Gerasimenko
and even put a lander on this few-kilometre-diameter
object, a technological masterpiece \citep{2019Filacchione,2020Filacchione}. Comets obviously
formed in the outer Solar System, as they are icy. Still, the
Stardust mission also brought back ‘hot minerals’ from
comet 81P/Wild 2, indicating some transport of matter
from close to the Sun to the comet-forming regions \citep{Brownlee2006a}.
So, in general, comets preserve some features of their
birth. They also break apart easily - that is, they have a
low tensile strength - and are little altered, indicating
a slow contraction of pebble- size dust and ice aggregates
during their formation \citep{Blum2017}.
By contrast, asteroids, which are found closer to the
Sun, are rather dry, and the smaller ones are rubble piles
that are heavily bombarded, as seen by recent missions
like Hayabusa \citep{Haffert2019} to asteroid 162173 Ryugu \citep{2019Watanabe} or OSIRIS-REx
to asteroid 101955 Bennu \citep{Barnouin2019}. Besides implying that the
story of planet formation involves many collisions, asteroids
are also parents to chondrites, a class of meteorites \citep{Weiss2013}.
Chondrites appear to be little altered since their formation
some 4.5 billion years ago, sharing composition and
age with the Sun. They also include millimetre-sized
chondrules (giving this meteorite class its name), spherical
particles that have no rival on Earth. Chondrules set
a size scale and are thought to be formed in a gas-rich
environment by rapid melting and rapid cooling of dust
aggregates \citep{2003Scott,2011Sears}. If chondrules collide when hot, they might
form what is known as compound chondrules. The condition
that the chondrules collide when still hot requires a
certain number density. Thus, the presence of compound
chondrules, which seems to be a detail, gives insight into
particle densities of the solar nebula as the protoplanetary
disk that formed the Solar System \citep{Bischoff2019}.
In addition, the composition of planets tells a story of
their formation. Gas giants on the one hand and rocky
planets on the other are signposts of the ingredients that
went into their formation selectively - gas and dust. As
one specific example of what can be learned beyond this
basic information, consider Mercury. All terrestrial planets
are thought to have an iron core, because the heavy
elements (and iron is most abundant) sediment to the
centre as the planet heats up due to gravitational energy
during formation and radioactive decay. However,
Mercury is especially rich in iron. An obvious mechanism
suggested to explain this finding has been mantle
stripping by a planet-sized impact \citep{Benz1988,Helffrich2019,Stewart2013}. But this mechanism
might not be the ultimate answer, as it should, in
general, deplete volatile elements, which conflicts with
findings by the MESSENGER mission \citep{Peplowski2011}. Furthermore,
there is a general trend that planetary iron content
decreases with distance from the Sun, giving hints on
other processes \citep{Mcdonough2020} that might have occurred during the
early phases of planet formation.
There are many more details of planets that might
give hints on their formation. Planets surrounding stars
others than the Sun are a rich source of information, and
much progress has been made since the first detection of
an exoplanet in 1995 (\cite{Mayor1995}), although we do not discuss
it in this Review.
Besides all the tremendous work done on planets,
the evolution from dust to planets in its protoplanetary
disk phase takes a few million years and cannot be
unravelled from a look at the final product alone. The
Atacama Large Millimeter Array (ALMA) provides a
powerful tool for the analysis of the birth place of planets,
the so-called protoplanetary disks. A large variety of
disk sizes, disk masses and dust distributions have been
detected \citep{Hendler2020,Trapman2020}. In addition, detailed structures such as rings
or spiral arms have been detected in several systems;
these structures are often attributed to planets forming
in these disks \citep{Andrews2018}. With the SPHERE (Spectro- Polarimetric
High-contrast Exoplanet Research) instrument at the
Very Large Telescope (VLT), another powerful tool for
observations of protoplanetary disks and embedded
planets has come into play. Because this instrument
works in the visible light spectrum and the stellar light is
blocked, the instrument is sensitive to dust and to planets
within the system. The state of the art of this discipline
is observation of systems such as PDS 70 (Fig. \ref{fig:1}), where
planets are directly observed in a protoplanetary disk \citep{Keppler2018,Haffert2019}.

\section{Planet formation models} \label{sec:models}
\begin{figure*}[ht!]
\plotone{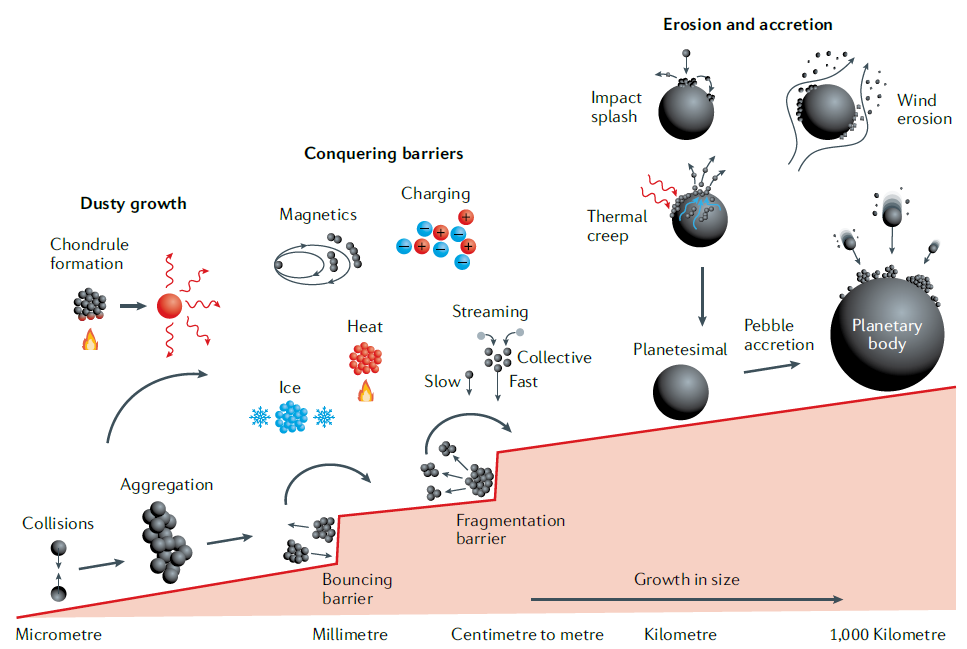}
\caption{\small{{Processes important during evolution from dust to planetary bodies that are accessible to laboratory and
microgravity experiments.} Evolution proceeds from left to right. Critical barriers in growth are depicted as steps, and
mechanisms to conquer them are positioned above. Critical ingredients also include chondrule formation and destructive
or morphology- changing processes for planetesimals.}
\label{fig:2}}
\end{figure*}
\noindent
In this section, we sketch out how planets are thought to
form, before revisiting the topic in greater detail in later
sections. Planets form in protoplanetary disks simultaneously
with the central star. In general, protoplanetary
disks are a mixture of dust and gas, with about 1\% of
the mass being in the solid phase \citep{Hayashi1985}. Dust evolution,
including planet formation processes, changes this ratio
continuously \citep{Sanchis2021}. Observations of dust emission reveal that
disks typically have lifetimes of a few million years \citep{Haisch2001}.
Typical disk masses given are a few per cent of the stellar
mass, but this figure is highly variable, and the mass also
decreases over time as viscous evolution and internal
and external photoevaporation drain the disk \citep{Schib2021,Hutchison2021,Haworth2021}.
There are several barriers to planet-sized bodies
forming out of the dust in protoplanetary disks (Fig. \ref{fig:2}),
and planet formation models endeavour to explain how
these barriers are overcome. Current models describe
terrestrial planet formation as a multi-step process. This
process starts with micrometre-size solid grains. Such
small grains have small relative velocities, driven by
Brownian motion, which nevertheless lead to collisions,
sparking the first growth phases via sticking collisions \citep{Weidenschilling1993}.
For bodies up to kilometre size, called planetesimals,
self- gravity is not important. Therefore, these growth
phases offer many open questions. For example, how
strong are the sticking forces for materials ranging from
cold ice to hot silicates? What are the limits of aggregation
through collisional sticking before bouncing or
fragmentation prevails? How efficient are hydrodynamic
instabilities in concentrating pebble-sized solids to the
point of gravitational collapse? The combined efforts of
experimental studies and numerical simulations have
formed the detailed picture of planetesimal formation
that is now known \citep{Johansen2014}.

Figure \ref{fig:2} gives an overview of the
different aspects that are important for planet formation,
and the experiments that have been carried out, as discussed
below. Much is determined by gas–grain interactions
for small bodies, and Appendix \ref{ch:particlemotion} details the dynamics of
solids governed by gas drag. Appendix \ref{ch:simulations} describes numerical
simulations that are used.
Numerical simulations predict fast growth by hit-and-
stick collisions - ‘fast’ meaning that millimetresized
aggregates in the inner few astronomical units
(AU) of a disk form on timescales of only 1,000 years \citep{Dominik1997,Wada2009,Wada2013,Okuzumi2012}.
Ground- based experiments have confirmed this fast
growth in two dimensions (2D) \citep{Wurm1998,Wurm2000}, but free collisions
in three dimensions (3D) are not feasible to study in
ground-based experiments. Collisional growth from
micrometre- sized dust grains to larger agglomerates
requires microgravity experiments for realistic
experiment conditions.
As the dust agglomerates grow in size, the relative
velocities are no longer driven by Brownian motion.
Rather, vertical settling in the disk and radial drift
towards the star determine the relative velocities between
solids. For agglomerates in the millimetre size range,
the relative velocities are still small. Experiments have
shown that collisions in this regime lead to compaction
of the agglomerates \citep{Weidling2009}. This compaction means that when
millimetre-sized agglomerates collide at small velocities,
they bounce off each other instead of sticking together.
This has been shown in numerous ground-based
experiments in a 2D setup \citep{Kelling2009,Kruss2016,Demirci2017} and was also confirmed
in free collisions in microgravity (discussed below).
Because particle growth comes to a natural halt in this
regime, it has been described in the literature as the
bouncing barrier \citep{Zsom2010}. Collision experiments provide a data
base for the outcome of an individual collision depending
on a variety of input parameters including size,
collision velocity and composition. Based on all these
possible collisional outcomes, numerical simulations can
then follow the evolution of solids from individual dust
grains to larger aggregates in specific locations for all
kinds of models of protoplanetary disks 
\citep{Andrews2018, Drazkowska2021, Pinilla2017}.
This first growth barrier is already a puzzle to be
solved by models. But even for dust agglomerates that
did grow further, there are additional barriers to be
overcome. If at least some dust agglomerates were larger,
these ‘lucky winners’ could grow further by collecting
mass from smaller particles \citep{Windmark2012}. Such mass transfer has
also been observed in ground- based experiments \citep{Teiser2009a,Teiser2009b,Meisner2012,Meisner2013}.
However, it remains to be explained how the lucky winners
formed in the first place. Additionally, growing
dust agglomerates decouple from the gas but are still
efficiently decelerated by gas drag. This drag leads to a
strong radial drift, which can be of the order of 1 AU per
100 years for metre-sized objects \citep{Weidenschilling1993,Weidenschilling1977}. For larger objects,
the radial drift is slower (Appendix \ref{ch:particlemotion}), so the lifetime of objects
beyond the metre size increases again with the size of
the objects. In the literature, this size- velocity regime
is also called the metre-size barrier or the drift barrier.
Growth beyond this barrier therefore has to be fast
enough that the material is not lost to the central star.

\noindent
Additionally, the relative velocities become large, so
that erosion and catastrophic fragmentation become
dominating processes.
Because collisions alone cannot lead to planet formation,
current models include hydrodynamic processes
in the protoplanetary disk, such as streaming
instabilities \citep{Johansen2007,Chiang2010}, which can concentrate particles of a
certain size, so- called pebbles. The size of these pebbles
is defined by a critical Stokes number $(St \approx 1)$, the ratio
between the gas–grain coupling time and the orbital
period. In absolute numbers, pebble size might be anything
from centimetre to metre scale \citep{Johansen2014}. If the particle
concentration gets larger than a critical density, gravitational
attraction between the solids then leads to the
direct growth of planetesimals from a dense cloud of
pebbles. Although these models describe a way to conquer
various barriers, they still require a minimum size of
the dust agglomerates to exist. Thus, the formation
of these pebbles still has to be explained. Microgravity
experiments have provided new ideas about how to
proceed beyond the classical bouncing barrier.
Several mechanisms exist that could lead to the
critical agglomerate size being reached or exceeded.

\noindent
As shown in Appendix  \ref{ch:protodisks}, the temperatures in protoplanetary
disks decline with increasing distance to the central star
so that volatiles such as water ice, carbon dioxide or
organic materials condense at characteristic distances
from the star, often called icelines or snowlines. This
condensation changes the density profile of the protoplanetary
disk, so that pressure excursions occur at
the different icelines, which can locally stop the radial
drift \citep{Dullemond2018,Pinilla2017} and therefore make coagulation easier. Selective
condensation of vapour close to the corresponding iceline
also can accelerate particle growth \citep{Ros2019}. Additionally,
the collision properties of the grains change with the
adsorbates on their surface, directly influencing the collision
dynamics. The water iceline, for example, has been
treated as guarantor for collisional growth by various
models \citep{Dominik1997,Kataoka2013,Saito2011}. Experimental results are needed in this
parameter regime, as the material properties (such as
surface energy) determine whether these models can be
applied or not.
Once planetesimals have formed, they are still subject
to a severe gas drag and to impacts of small particles.
Most theoretical models do not include erosion
processes, either by wind or by impacts. However,
experiments provide suitable means to study how stable
planetesimals are against potentially destructive
processes. Because planetesimals have only very small
gravitational acceleration on their surface, microgravity
experiments are needed to determine the influence of
erosive processes.
Collisions become important again for further evolution
of planetesimals. Pebbles are currently considered to
be of critical importance for further evolution. Although
they might not grow themselves in mutual collisions
(because of the bouncing barrier), they might still provide
a vast reservoir of material. In particular, because
pebbles efficiently drift radially, they can be accreted by
planetesimals and their descendent protoplanets \citep{Lammer2021,Dash2020,Ndugu2021,Lenz2020}.
Planetesimal collisions among each other and with
growing protoplanets might also be of importance \citep{Brugger2020}. Full
collisions simulating these phases and the last phases
of large impacts cannot be investigated by laboratory
experiments because the bodies involved are too large,
so numerical simulations are the only method to study
these processes. During these late phases, the protoplanetary
disk disperses to become a debris disk \citep{Watt2021,Krivov2021}, but these
are beyond the scope of this Review. More details can be
found in \citet{Armitage2009}.\\

\section{Microgravity platforms} \label{sec:microgravity}\noindent
To exclude the influence of Earth’s gravity field, different
methods can be applied to laboratory experiments,
ranging from free fall, in different environments and
for different timescales, to levitation techniques on the
ground. Levitated particles are not completely free, but
a certain degree of motion (for instance, 2D) might be
allowed. One example of levitation techniques currently
used is acoustic traps, in which particles are captured
at the nodes of an ultrasonic acoustic wave induced by
an array of loudspeakers \citep{Lee2018}. Another method is levitation
by thermally induced gas flow in porous aggregates \citep{Kelling2009}.
This method is suitable for the levitation of porous
particles and allows study of slow particle collisions in
2D relevant for planet formation.

\noindent
This method has been
used extensively to study growth up to the bouncing
barrier and its dependence on external parameters such
as temperature \citep{Demirci2019a} or particle properties like grain size \citep{Kruss2016},
magnetism \citep{Kruss2018,Kruss2020} and composition \citep{Demirci2017}.
For the study of many aspects of planet formation,
levitation techniques are not suitable, either because free
particle movement in 3D is needed (as in many collision
experiments) or because the properties of the studied
sample depend on the residual gravity - for instance,
regolith (a mixture of sand and dust) on asteroid surfaces.
In these cases, experiments require free-fall conditions.
This can be achieved with numerous microgravity
platforms, which offer free- fall conditions of different
timescales and different levels of residual gravity. Table 1
shows an overview of different microgravity platforms,
as used for research on planet formation. Often, only
a short period of microgravity is needed, for instance
for experiments to study individual, slow collisions.
In this case, a small drop facility with a drop height of a
few metres might be sufficient. Some groups work with
stationary vacuum tubes, in which the particles are in
free fall, while the observation apparatus is dropped
simultaneously \citep{Waitukaitis2014,Blum2014}. Other groups use a compact experiment
rack that is dropped all together, including the
whole experimental setup \citep{Sunday2016,Murdoch2017}. In addition, experiments
with reduced but non- zero gravity are possible with
these facilities, so that the gravitational attraction on
smaller planetary bodies can also be simulated.
Larger research facilities such as the drop tower
at Bremen (ZARM) offer 4.7\,s of microgravity if the
experiment is only dropped or 9.2\,s if the experiment
is launched by a catapult. In both cases, a gravity level
of $10^{-6}$\,g $(g = 9.81\,m\,s^{-2})$ is reached \citep{VonKampen2006}. Other drop towers
also exist, which offer microgravity for about 4 s, such
as the Beijing drop tower \citep{Liu2016}. However, to our knowledge,
the Bremen drop tower is the only one that has been
used in the context of planet formation.
Several of the experiments discussed below were
performed on parabolic flights. This kind of platform
offers microgravity of 20–22\,s duration (per parabola)
with a residual acceleration smaller than 0.05\,g (\citep{Pletser2015}).
Owing to the significant level of residual acceleration,
this platform is not suitable for all experiments.
However, this platform also offers experiments under
partial microgravity, either by flying a different trajectory
with the aircraft or by using a linear stage \citep{Guettler2013} or a
centrifuge \citep{deBeule2013,Musiolik2018,Demirci2019b}.
The microgravity platforms of drop towers and parabolic
flights are also frequently used to perform collision
experiments on granular matter that are not focused on
but strongly related to planet formation, such as studying
the clustering transition from a gas- like to a solid- like
state \citep{Opsomer2012,Sack2013,Noirhomme2018,Harth2018}.
Experiments with longer duration can be performed
in suborbital flight vehicles, for example on traditional
sounding rockets or on similar platforms that are currently
emerging and are also intended to be crewed.
Timescales of 1–6\,min are possible, depending on
the platform. Such platforms have been used by various
teams to study particle collisions and growth \citep{Goldhirsch1993,Falcon1999,Blum1999b,Harth2013,Brisset2016,Brisset2017a,Colwell2016,Yu2020}.
Depending on the height of the apogeum and the vehicle’s
avionics, residual acceleration of $10^{-3}$\,g to $10^{-4}$\,g can
be reached.
An even longer experiment duration can only be
reached in orbit, with various platforms available. The
International Space Station (ISS) offers the longest
experiment duration, as the experiment time, in principle,
is not limited. Experiments on the ISS have been
conducted within the framework of student research
with strong limitations on space, mass and power
consumption \citep{Brisset2019,Steinpilz2020b}, but other experiments on a more
sophisticated level are planned \citep{Aumaitre2018}. More complex experiments
have been carried out on the Space Shuttle \citep{Colwell2003,Blum1999a,Blum2002}.
In this case, the maximum operation time is limited
to the corresponding mission. An additional opportunity
for experiments in orbit is given by small satellites
(CubeSats), which can be deployed from the ISS or as
secondary payload from large launchers (in combination
with the primary payload). Their operation time is
limited by the orbit decay of the satellites, on the order
of months. Experiments on small satellites are currently
also planned \citep{Lightholder2017,Jarmak2019}.

\section{Laboratory experiments to fill the gaps} \label{sec:experiments}
\subsection*{Colliding}\noindent
The main drivers for early (and late) planet formation
are collisions. For laboratory experiments, gravity is a
real ‘downer’ here as it simply limits the time to observe
free, slow individual collisions or the growth of a cloud
over a longer timescale.\\

\noindent
\textit{\textbf{Very early growth.}} Planet formation starts with collisions
of micrometre- sized dust grains. Some of these
grains formed in old stars or during supernovae and
survived passage through interstellar space and into
the protoplanetary disk. Other grains might have come
from material that was destroyed by shocks or high
temperatures and only re- condenses within the disk \citep{Tscharnuter2009}.
These grains are small and couple very well to the gas.
Therefore, their transport in the disk is initially dominated
by Brownian motion, as are the collisions. At the
given grain size (or mass) and temperatures, they
move with millimetre per second speeds through the
gas $(1/2mv^2 = 1/2k_BT$, where $m$ is grain mass, $v$ is grain
velocity, $k_B$ is Boltzmann’s constant and $T$ is temperature).

\begin{figure}[]
\epsscale{1.2}
\caption*{\small{\textbf{Table 1 | Overview of microgravity platforms}}}
\plotone{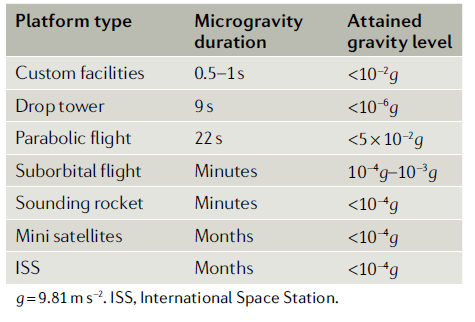}
\end{figure}
\noindent
However, the motion is diffusive, and particles
rapidly change their direction. For particle growth, such
direction changes are important because a single grain
approaching a large aggregate might not pass the outer
perimeter of that aggregate before changing direction
and eventually making contact on the outside rather
than within the interior of the aggregate.
Growth in these early times is known as hit- and- stick,
because grains simply stick in a rigid way upon first contact.
One might think that at a velocity scale of millimetres
per second, there should be plenty of time for
experimental observations even in a ground- based setting.
However, gravity is still acting, and at pressures of
millibars or lower - mandatory for planet formation
experiments - grains settle quickly. Aggregates in this
early phase are characterized by a fractal dimension $d$,
where the mass $m$ of an aggregate scales with aggregate
radius as $m \sim r^d$ with $d \leq 3$. Aggregates that are grown by
adding individual particles grain by grain to a cluster are
known as particle–cluster aggregates, and although they
become very porous, they have $d = 3$. If collisions always
happen between clusters of similar sizes, the aggregates
are known as cluster–cluster aggregates, and $d < 2$. The
latter process is assumed to take place self- consistently
in a cloud of particles \citep{Wurm1998}.
The way that grains grow is important as it determines
the gas–grain coupling time $\tau$ (Appendix \ref{ch:particlemotion}). Aggregates
with $d < 2$ do not change their coupling time during
growth because the mass–surface area ratio remains
constant, and they continue to move and collide slowly.
For $d = 3$ growth, $\tau \sim r$, and larger aggregates behave differently
from smaller ones, for instance settling faster in
the protoplanetary disk. This difference sets the stage
for all further particle evolution. An extreme model \citep{Kataoka2013}
assumes aggregates to grow with low fractal dimension
up to kilometre size. Such growth might depend on the
sticking properties of the grains, and the model assumes
that very small water ice grains (100\,nm) are much stickier
than silicates (see below for details on ice). But it
would circumvent some of the problems encountered,
such as the drift or fragmentation barriers.
Brownian motion and its ballistic part can be studied
in drop tower experiments \citep{Krause2004,Blum1996}. An important work was
an experiment onboard the Space Shuttle in 1998 to see
aggregation beyond single collisions \citep{Blum2000b}. Interestingly, the
fractal dimension of the aggregates was rather low -
that is, the aggregates were chain- like. This finding can
be explained by the fact that aggregate rotation is also a
diffusive (Brownian) motion \citep{Paszun2006}.\\

\noindent
\textit{\textbf{Compaction phase.}} Simulations and experiments indicate
that aggregates restructure if they reach approximately
millimetre size. As the aggregate mass increases,
the kinetic energy increases. This energy can no longer
be dissipated in hit- and- stick collisions. Energy is
instead dissipated as grains within an aggregate inelastically
roll about their contacts or slide along each
other \citep{Dominik1997}. Experiments are essentially found to be in
agreement with these simulations. For instance, in drop
tower experiments, restructuring is observed, depending
on the collision velocity \citep{Blum2000b}. As a consequence of
restructuring, aggregates become more and more
compact \citep{Weidling2009}.

\begin{figure}[]
\epsscale{1.2}
\plotone{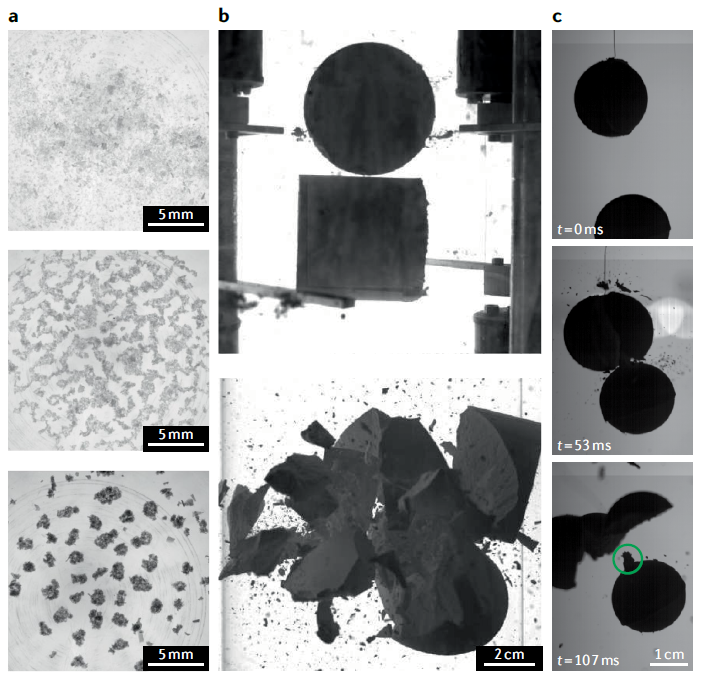}
\caption{\small{{Examples of different collision outcomes.} \textbf{a|} Growth from small (10$\,\mu m$)
agglomerates to millimetre-scale particles at the bouncing barrier. The dust particles
were levitated by placing them over a hot plate at low ambient pressure, thereby
allowing them to move freely in two dimensions. \textbf{b|} Collision leading to total
fragmentation. \textbf{c|} Collision with mass transfer. One agglomerate is destroyed, while the
other grows further. The collisions in parts \textbf{b} and \textbf{c} are performed in a custom drop
facility, with the two collision partners released with a time lag $\Delta t$, giving a relative
velocity between the bodies of $g\Delta t$, where $g = 9.81$\,ms$^{-2}$. Part \textbf{a} Credit: Kruss, Teiser
\& Wurm, A\&A, 600, 2017, reproduced with permission \textcopyright ESO. Part \textbf{b} adapted with
permission from \citep{Schraepler2012}, IOP. Part \textbf{c} adapted with permission from \citep{Beitz2011}, IOP.}
\label{fig:3}}
\end{figure}

Compact in this context means volume filling
fractions of 30–40\% \citep{Meisner2013,Teiser2011b}. Eventually, efficient restructuring
is no longer possible and can no longer provide
enough energy dissipation, and collisions become more
and more elastic. Note that for silicates a long fractal
growth phase is unlikely.\\

\noindent
\textit{\textbf{The bouncing barrier.}} There have been a number of
microgravity experiments testing how large these aggregates
might grow in further slow collisions \citep{Brisset2016,Brisset2017,Kothe2013,Weidling2012,Weidling2015,Langkowski2008}.
Some of these collisions lead to sticking at velocities
below 1\,m\,s$^{-1}$, which are relevant for protoplanetary
disks. However, not all collisions are sticky, and aggregates
often simply bounce off each other. As in protoplanetary
disks, dust aggregates speed up again (couple
to the gas) before they collide the next time, and even a
low coefficient of restitution does not warrant sticking
after some collisions. We note that this is in contrast to
collisions in Saturn’s rings, for example, a case in which
particles ‘remember’ their collisional history. Therefore,
the existence of some sticking events does not prove
that growth proceeds ever further. On the contrary,
even if aggregates stick to aggregates, the connections
are so weak that further collisions can dissolve a cluster
of aggregates again into its constituent aggregates.
Therefore, there is a phase in early pre- planetary evolution
in which bouncing prevails, known as the bouncing
barrier \citep{Zsom2010}.
This halt in growth is quite severe. The exact size at
which bouncing dominates collisions depends on sticking
properties and collision velocities. Both depend on
the specific model of the protoplanetary disk and the
location of the aggregates within it. Nevertheless, putting
a number to it, growth gets stalled at the millimetre
size. So far, there is no self- consistent microgravity
experiment on the bouncing barrier, owing to limited
timescales and the requirement that the restitution
must somehow be balanced after each collision to prevent
collisional damping and slow- down of collisions.
Therefore, so far only levitation experiments on the
ground have studied the self- consistent evolution of
an ensemble of particles at the bouncing barrier. In a
2D arrangement, particles grow until they reach the
bouncing barrier. An example of the growth of small
(10\,\textmu m) particles to millimetre size and into the bouncing
barrier is shown in Fig. \ref{fig:3}\textbf{a}. Further sticking is easily
undone by subsequent collisions because connections
in clusters are weak \citep{Kruss2016,Demirci2017,Demirci2019a,Kruss2017,Kelling2014,Jankowski2012}. Therefore, the growth
of agglomerates into a bouncing barrier should be
considered as a robust result.\\

\noindent
\textit{\textbf{Fragmentation.}} Aggregates in protoplanetary disks
do not easily grow further. However, planets exist, so
evidently assemblies of particles larger than millimetre
scale must be possible. One can therefore ask how collisions
change if dust aggregates of centimetre or tens of
centimetre size are present, whatever their origin might
be. Microgravity experiments are mandatory to answer
this question, because gentle collisions between such
objects cannot be studied on the ground. There have
been a number of such experiments \citep{Meisner2012,Beitz2011,Deckers2013,Deckers2014,Kothe2010,Schraepler2012,Katsuragi2018}. All these
studies indicate that collisions between similar- sized
(similar- mass) objects at velocities smaller than 1\,m\,s$^{-1}$
do not lead to sticking. Only if collisions are super- slow,
far below applicability to protoplanetary disks, has sticking
been observed. There is instead a growth barrier due
to fragmentation. This occurs because - thanks to the
larger mass - the energy at increasing velocities eventually
becomes large enough to destroy contacts again and
consequently destroy the whole aggregates. An example
of a catastrophic collision between two equally sized
aggregates is shown in \ref{fig:3}\textbf{b}. In the case that agglomerates
grow beyond the bouncing barrier, the fragmentation
barrier will stop further growth at particle sizes
that are only slightly larger than the sizes resulting from
the bouncing barrier.
Dust in protoplanetary disks is observed for disks up
to a few million years old according to the spectra of
their central stars. If agglomerate growth stopped at a
certain stage, there should be none, as it would have been
lost due to radial drift and accretion to the central star.
Therefore, fragmentation might be important for producing
the observed dust \citep{Husmann2016}, although other mechanisms
might also be at work \citep{deBeule2013dust} (see the discussion of thermal
effects below).\\

\noindent
\textit{\textbf{Mass transfer.}} Fragmentation also occurs if the collision
partners are different, either in size or in volume
filling. There is a subtle but important difference,
though, from the case of equal- sized aggregates: the
larger (or more compact) of two aggregates can remain
intact while the smaller (or more porous) one is
shattered \citep{Teiser2009a,Teiser2009b,Teiser2011b,Beitz2011,Deckers2014,Wurm2005,Teiser2011a}. In addition, a fraction of the
smaller aggregate adheres to the large aggregate, which
thus grows again. Figure \ref{fig:3}\textbf{c} shows an example of a collision
with mass transfer \citep{Beitz2011}. This process is a hybrid of
destruction and growth. Whereas the smaller collision
partner is destroyed, dissipating kinetic energy, part of
its mass feeds the larger collision partner. Even up to
high velocities of tens of metres per second, the larger
body moving through the protoplanetary disk can sweep
up material of small impactors in this manner \citep{Meisner2013,Schraepler2018}. Freefall
experiments show that this result is not an artefact
of gravity \citep{Beitz2011}.
It has been shown that planetesimals can indeed form
this way if some larger seed particles sweep up the small
particles \citep{Windmark2012}. The fact that the bouncing barrier prevents
many particles from becoming larger is beneficial in this
context. The timescales for growth seem long, so it might
be questionable whether this mechanism is a primary
mechanism for planetesimal formation. However, the
process itself is of importance in all collisions occurring
at higher speeds.

\subsection*{How to conquer barriers}\noindent
Particle motion in dense clouds can lead to concentrations
and planetesimal formation eventually taking
over from collisional growth \citep{Johansen2014}. However, as noted
above, these processes need particles of a critical Stokes
number, which corresponds to particles with sizes of
centimetres or tens of centimetres. Regular growth that
depends on simple surface forces does not give access
to this size range, because the bouncing and fragmentation
barriers prevent aggregates from growing large
enough.

\begin{figure}[]
\epsscale{1.2}
\plotone{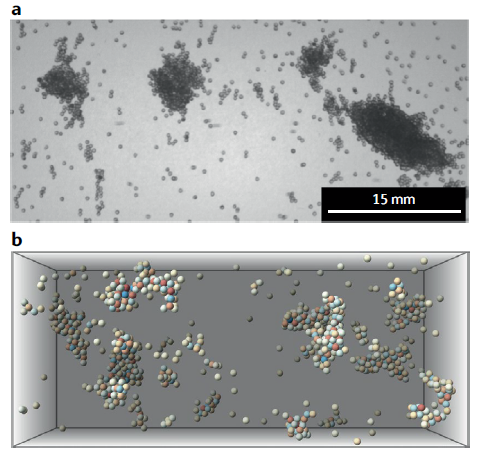}
\caption{\small{{Observed and simulated clusters of charged
grains.} \textbf{a|} Observations. \textbf{b|} Simulations. The charge of
grains might enable them to grow from the regular
bouncing barrier (at millimetre scale) into drag instabilities,
which exist for centimetre or larger particles. The
experimental observations are made in a drop tower.
Figure adapted from \citep{Steinpilz2019}, Springer Nature Limited.}
\label{fig:4}}
\end{figure}

There is evidence from aggregate chondrules
in meteorites that a further growth phase occurred \citep{Simon2018}.
Thus, a number of ‘supporting actors’ have been evaluated
to shift the bouncing and fragmentation barriers
to larger sizes; much of this work is emerging
from laboratory experiments, including microgravity
experiments.\\

\noindent
\textit{\textbf{Charged particles.}} One of the most promising processes
is electric charging of grains through bouncing collisions.
After several collisions, this charging can provide
strong attractive forces. In a number of microgravity
experiments, charging between grains was studied and
the aggregation of clusters was observed \citep{Lee2018,Waitukaitis2014,Steinpilz2020b,Steinpilz2019,Steinpilz2019a,Jungmann2018,Love2014,Marshall2005}. More
recently, the formation of clusters of several centimetres
in size due to electrostatic forces was shown explicitly,
through drop tower experiments and numerical
simulations \citep{Steinpilz2019a} (Fig. \ref{fig:4}). Even the charge- driven growth of
large (5\,cm) agglomerates from small (165\,\textmu m) solid particles
has been observed directly \citep{Teiser2021}. Thus, the bouncing
barrier might in fact be the beginning of a new growth
phase as particles are charged in collisions and attractive
forces are increased in comparison to typical surface
forces acting. At present, charged aggregation is the only
mechanism proven to span a wide range in particle size.
Other processes might trigger the formation of larger
aggregates in certain disk locations.\\

\noindent
\textit{\textbf{Magnetics.}} If electrostatics enable aggregation, a natural
question is whether magnetic fields and magnetic
dipole attractions could also be an effective ‘glue’. This
effectiveness would have to be compared with electric
dipole interactions. In fact, the net electrical charge
on grains might not be dominant for stability compared
with dipoles and higher- order charge moments
on grain surfaces \citep{Steinpilz2020b}. Therefore, attraction should also
work for magnetic particles. There are two possibilities:
either particles are permanent magnets, or they contain
ferromagnetic grains that are magnetized by magnetic
fields within the protoplanetary disk. The former case
has been studied in ground- based experiments, but also
under microgravity and in numerical simulations \citep{Nuth1994,Dominik2002,Nuebold2002b,Nuebold2003,Yu2019,Opsomer2020}.
In the case of permanent magnets, aggregation is accelerated.
Because dipoles are not necessarily aligned,
large 3D structures can form readily, but it is uncertain
whether permanent magnets are abundant in
protoplanetary disks.
Ferromagnetic iron becomes magnetized in the fields
present in protoplanetary disks. Such magnetization also
promotes aggregation \citep{Kruss2018,Hubbard2014,Kruss2020b}. Because of the alignment
of the dipoles, chain- like clusters of aggregates form in
this case. These chains are still larger than clusters of
non- magnetic material, and this puts a bias on growth.
Because magnetic fields are stronger closer to a star,
aggregates grow preferentially closer to the star, as long
as the temperature is not beyond the Curie temperature.
These results from experiments might offer an
explanation of the general trend in the Solar System
that iron- rich bodies are found further inwards, with
Mercury being the most iron-rich \citep{Mcdonough2020}.\\

\noindent
\textit{\textbf{High temperature.}} Temperatures in protoplanetary
disks vary from only a few kelvin, far from the star,
to sublimation of essentially all solids at 2,000 K close to
the star. The local temperature changes the local composition
of solids, grain sizes and viscosity - all ingredients
for sticking. Laboratory work is rare on the
high- temperature side. There are levitation experiments
showing that the particle size resulting from the bouncing
barrier shifts for tempered silicate and iron- bearing
dust \citep{Demirci2017} - that is, the resulting final size decreases for temperatures
larger than 1,000 K. One origin to which this
outcome can be traced is the variations in composition \citep{Bogdan2020}.
Viscosity changes, which might be important for warm
dust, have also been probed in levitation experiments \citep{Demirci2019b}
and in classical ground experiments \citep{Bogdan2019}. These experiments
imply somewhat increased capabilities of sticking
at higher temperatures. Overall, there is no final, general
statement on the influence of temperature on growth yet.

\noindent
\textit{\textbf{Chondrules.}} The need for levitation and microgravity
experiments emerges in the context of high temperatures
because molten or very ductile grains on surfaces are not
convenient for experiments. This is especially important
in the field of chondrules, the millimetre- sized spherical
grains found in meteorites. It may or may not be by
chance that these abundant spherical particles are similar
in size to the particles generated by the bouncing
barrier. In any case, they were molten and cooled rapidly
in the Solar Nebula. Some of them even collided while
still hot and formed compound chondrules \citep{2017Bischoff}. Collision
experiments on the ground can be carried out as long as
the particles are not yet molten \citep{Bogdan2019}. These experiments
show a decrease of the coefficient of restitution beyond
1,000 K for basalt particles, again hinting at changes
in particle growth with temperature. Experiments in
short- time microgravity have also been carried out with
chondrules, especially with respect to understanding fast
melting and cooling \citep{Nagashima2006,Poppe2010,Guettler2008}. Further microgravity experiments
and gas- flow levitation produced and used dustcovered
chondrule analogues. One motivation here is to
answer whether larger aggregates can grow from such
dust- covered chondrules \citep{Beitz2012,Beitz2013}. Indeed, sticking velocities
change and would probably allow the formation of small
dust- glued chondrule clusters.
In all these cases, there is a shift from ground- based
levitated to microgravity experiments motivated by the
need to have contact- free, slow collisions or cooling
undisturbed by gravity- induced effects such as free convection.
While the cooling experiments elaborate more
on the possible mechanisms for chondrule formation,
such as shocks, lightning or other processes, the collision
experiments show that temperature matters for sticking.
Hot collisions do not directly conquer the bouncing barrier;
but if only small factors are needed to move from
collisional growth to hydrodynamic concentration, these
subtle details might be the trigger.

\noindent
\textit{\textbf{Low temperatures and ices.}} The importance of subtle
details also holds for cooler conditions. Much potential
to form larger bodies has been attributed to the cold
regions of protoplanetary disks. Composition- wise,
what condenses in such regions is ice of different types.
At about 150 K at protoplanetary disk pressures, water
changes from gas to ice. Because water is abundant, the
respective water ‘snowline’ at a few AU is one of the most
important icelines as it approximately doubles the local
mass of solids and changes the sticking properties of
aggregates. Everyday experience shows that snowballs
are very sticky, and indeed ice has been a favourite of collisional
growth research for years \citep{Hatzes1988,Bridges2001,Drazkowska2017}. More specifically,
the surface energy of water ice close to the melting point
is $\gamma \approx 0.1 - 0.2$\,Jm$^{-2}$  \citep{Gundlach2011,Aumatell2014}. This value is a factor of
10 larger than measured for silica \citep{Heim1999}. Additionally, collision
velocities below which micrometre- sized ice grains
stick have been found to be an order of magnitude
higher than for silicates \citep{Gundlach2015}.
Because the increased growth capabilities of water
ice had already been treated as a drive for planetesimal
formation, it came as a surprise that at reduced temperatures
the stickiness of water ice decreases greatly \citep{Gundlach2018,Gaertner2017}.
In fact, laboratory experiments have shown that the surface
energy is constant from freezing point (273\,K) down
to 200\,K but then decreases by two orders of magnitude
down to 170\,K, relevant for protoplanetary disks \citep{Musiolik2019}.
Conversely, silicates were found to be more sticky - having
a surface energy increased by a factor of 10 - at the
dry conditions of protoplanetary disks \citep{Kimura2015,Steinpilz2019gamma}. Therefore,
collisions of water ice might not be more favourable for
growth than those of silicates. Nevertheless, ice collisions
occur, and with respect to mass transfer and restitution
these are still interesting. Ice collisions have been
studied especially in ground- based experiments \citep{Deckers2016,Arakawa2011,Shimaki2012a,Shimaki2012b,Yasui2014}.
In microgravity experiments, slower collisions have also
been observed \citep{Aumatell2014,Heisselmann2010,Hill2015a,Hill2015b,Aumatell2011}. All in all, water ice collisions
are important but might not provide a solution to the
bouncing barrier problem.
There are also other icelines and ices to collide. $\rm CO_2$
has been studied as one of the next most abundant ices,
but it has not been found to be stickier than water ice
or silicates in collision experiments \citep{Musiolik2016a,Musiolik2016b}, although work
on this topic is ongoing. Additionally, organics have
been studied experimentally in different temperature
ranges \citep{Kudo2002,Homma2019,Bischoff2020}. They might, in fact, have a sweet spot of
temperature to be sticky, but it is currently unknown
how these pure organics would interact with the other
materials.

\noindent
\textit{\textbf{Sublimation and condensation of ices.}} The story of water
ice is not finished by considering only collisions. Being
volatile, water ice is subject to ongoing sublimation and
recondensation around the snowline. These processes
have different effects. Sublimation by itself removes
water from all particles that drift inwards and cross the
snowline, an effect that still occurs in the present- day
Solar System on comets \citep{Gundlach2020,Fulle2020,Bischoff2019}.
In protoplanetary disks, if silicate dust grains within
mixed aggregates are set free, the size distribution of solids
at the snowline is changed. A reservoir of small particles
might be provided, upon which larger agglomerates
might feed and grow further. Particles on the outside
of the snowline might also directly grow by condensation
of water vapour as it diffuses outwards again \citep{Ros2019,Saito2011,Schoonenberg2017}.
Levitation experiments of sublimating water ice confirmed
the concept of break up into subunits that would
set free smaller dusty seeds \citep{Aumatell2011}. This is the only experimental
study we know of that dwells on this important
topic. Microgravity experiments are needed to explore
further, but such experiments might not occur in the
near future, because a long experimental duration is
needed, and volatiles are not easy to handle in space.
Additionally, sublimating ice increases the local gas
pressure in the disk, which changes the dynamics. At
the local pressure maximum, the inward drift of solids
is reduced or even stopped by an inverse pressure
gradient \citep{Dullemond2018,Pinilla2017}. Processes at the snowline therefore are a
viable mechanism to initiate planetesimal formation
locally.
Sublimation and recondensation or sintering can
also change the contacts between grains, making contacts
more stable and harder to break. On first sight, this
stability might look beneficial. However, it removes an
easy route for energy dissipation, and it might inhibit
collisional growth \citep{Sirono2017}. Experiments are needed to answer
these questions.

\subsection*{Particle transport}\noindent
Particle transport in protoplanetary disks is a rich field,
and we do not aim at a full review of all possible particle
motions here. Instead, we mention a small selection of
phenomena that are accessible to microgravity experiments:
radiation pressure, photophoresis and, in part,
collective motion.

\noindent
\textit{\textbf{Radiation pressure.}} Radiation pressure is a well known
concept in astrophysics and plays a vital role in highmass
star formation, for example \citep{Kuiper2010}. Radiation pressure
in a planetary context is most important for optically
thin systems, such as our Solar System or late planetary
systems that have little or no gas content, and in protoplanetary
disks in regions where light reaches dust particles.
These regions are mostly the inner edge, facing the
star, and the thin upper layers of the warped protoplanetary
disk. Radiation pressure is also important for all
dust grains set free by comets or as debris of collisions.
The ratio $\beta$ between radiation pressure force and stellar
gravity can be larger than 1 for micrometre grains.
These grains, known as $\beta$-meteoroids, are continuously
accelerated and kicked out of the system. For somewhat
larger grains, with $\beta < 1$, radiation pressure effectively
reduces the stellar gravity. Radial radiation pressure does
not change the orbit directly. However, the Poynting–
Robertson effect based on light aberration lets dust particles
spiral inwards to the star \citep{Wyatt1950,Klacka2010}. The dynamics can
be calculated for spherical particles. Measurements with
non- spherical particles are rare but have been carried out
for trapped graphite particles released to free fall for a
short time while a laser pulse hits the particle \citep{Krauss2004,Krauss2007}.
Radiation pressure is also capable of mixing dust
particles in the disks \citep{Vinkovic2009}. Such particle mixing, especially
in an outward direction, might explain how refractory
materials found their way into comets, even though
the materials condensed at high temperatures and comets
formed in the cold outer regions \citep{Zolensky2006}. Besides the two
experiments mentioned, and many technical exploitations
of radiation pressure such as optical tweezers,
radiation pressure is usually only calculated assuming
spherical grains. Non- spherical grains would be important
for inducing particle rotation, and in fact radiation
pressure might vary significantly for them, but this is
currently not a focus of experimental research.

\noindent
\textit{\textbf{Photophoresis.}} Whereas radiation pressure might be
more of a detail, photophoresis is not. In contrast to
radiation pressure, though, photophoresis is still not
a widely known process and is often ignored even if
radiation pressure is evaluated in the presence of gas
in protoplanetary disks \citep{Vinkovic2020}. This is the case even though
photophoresis is precisely a force acting on an illuminated
solid grain embedded in an optically thin gas. The
effects of photophoresis are similar to radiation pressure,
as light is absorbed by a particle and the particle
(usually) moves away from the light source. However,
whereas for radiation pressure the momentum is directly
transferred by the absorbed or scattered photons, photophoresis
is based on momentum transfer by gas molecules.
This process is far more efficient. In fact, if the
conditions are right, the photophoretic force in a gaseous
disk can be a million times stronger than radiation pressure
and can drive particles outward in protoplanetary
disks with ease \citep{Krauss2005,Wurm2006b}. Depending on the particle size and
thermal conductivity, particles are trapped in rings if the
disk is optically thin enough. This trapping can also be
part of a recycling process to keep dust in a ring close
to a star while most of the disk is already in the process
of dispersal \citep{Husmann2016}. A number of drop tower and parabolic
flight experiments have been carried out in which particle
motion on free, illuminated particles - dust and
chondrules - has been measured \citep{Wurm2010,Borstel2012,Loesche2013,Loesche2014}. The experiments
essentially quantified the photophoretic strength, which
depends on the thermal conductivity of the particles.
We emphasize that photophoresis by stellar illumination
only works in optically thin parts of the disk. Particle
transport therefore either occurs in the inner part (gap)
of the disc, or over the surface of the disk, albeit more
effectively \citep{Wurm2009f} than radiation pressure. However, thermal
radiation in a disk with temperature fluctuations or radiation
from a forming planet also induces photophoretic
motion, which can lift and trap particles and might be
important for chondrules being transported after they
form \citep{vanEymeren2012,Teiser2013,Loesche2016,Mcnally2017}.
Overall, photophoresis has been put on firm ground
in recent years by experiments, and analytical models
have been derived based on these experiments. In any
case, whenever particle motion in a gaseous disk is
considered with any kind of directed radiation being
present, photophoresis influences particle motion
strongly \citep{Krauss2005,Wurm2006b,Teiser2013,Loesche2016b,Cuello2016,Arakawa2019}.

\noindent
\textit{\textbf{Collective motion.}} What we term ‘collective motion’
here is a field that has gained much importance over
recent years \citep{Squire2018}. With bouncing and fragmentation, pure
collisional planet formation seems to have too many
fundamental problems to proceed all the way from
dust to planets. Therefore, the above- mentioned hydrodynamic
processes have emerged to bridge from sizes
of tens of centimetres to kilometres \citep{Johansen2007,Chiang2010}. So far, this has
exclusively been a domain of theory and numerical simulation,
apart from ‘real’ fluid experiments with liquids,
which are not the topic of this Review. Experiments that
directly approach a low- pressure atmosphere show
that a grain in a large group of grains in close proximity
behaves collectively, that is, reacts more sensitively to
disturbances, making it behave like a larger particle \citep{Schneider2019a,Schneider2019b,Capelo2019}.
The experiments give evidence that the effective gas–
grain coupling time $\tau$ (Appendix \ref{ch:particlemotion}) can be systematically
modified.
These results can be used to calculate corrections
to sedimentation or drift velocities depending
on the solid- to- gas ratio and closeness of particles, but
experiments remain far from simulating phenomena
such as streaming instabilities. It is currently not clear
how far solid/gas experiments might really penetrate
into this field, as the mix of gas and particle motions in
protoplanetary disks is complex.

\subsection*{Planetesimals are fragile}\noindent
One way or another, barriers or not, planetesimals do
form. But as their origin is dusty, they certainly do not
start off as pure consolidated rock. They are also small,
without much gravity to hold them together. This means
that young planetesimals are fragile objects, prone to
easy disassembly by the smallest disturbances. They
can only become planets if they survive any attempts at
erosion or destruction. Destructive mechanisms that are
important in this context are thermal creep, wind and
impacts.

\noindent
\textit{\textbf{Thermal creep.}} Like photophoresis, thermal creep is a
phenomenon that is not yet widely known or applied to
astrophysics. In fact, it is one of the topics that is driven
by laboratory and microgravity experiments. Thermal
creep and photophoresis are two sides of the same coin.
Whereas photophoresis describes the motion of a solid
embedded within a gas, thermal creep describes the
gas motion around a solid. From simple arguments of
momentum conservation, if a solid, heated by light on
one side, moves in the direction of the cold side, gas
must flow from the cold to the warm side. The effects of
thermal creep are mostly visible at low pressure, where
mean free path lengths of the gas molecules are comparable
to or larger than the length scales of a solid. If the
solid is the porous surface of a planetesimal, gas flows
within the pore space along temperature gradients.
A model has been constructed for thermal creep
within a dust sample based on ground- based measurements
of static pressures and mass flows in two
gas reservoirs connected by a dust tube \citep{Koester2017}. This model
was verified in microgravity experiments in which the
gas flow was traced directly by observing the motion
of small tracer grains coupled to it \citep{Kraemer2019,Steinpilz2017,Schywek2017}. In a setting
with a more complex temperature gradient or varying
dust properties, the gas pressure within the dust is no
longer constant. In particular, excess pressure can lead
to explosive ejection of dust aggregates from the surface
of a dusty body. This was nicely shown in drop tower
experiments \citep{deBeule2014}. A more detailed gravity dependence on
the particle ejections resulted from centrifuge experiments
on parabolic flights \citep{deBeule2013}, showing the transition to
zero g, for which only adhesion remains to keep a dust
pile together. Although dust ejections had been observed
before in experiments \citep{Wurm2006,Wurm2008,Kelling2011,Kocifaj2011}, it was unclear what the
origin was other than being related to low- pressure
physics.
The effect is now recognized as a major mechanism
for lifting dust on Mars \citep{Neakrase2016,Schmidt2017,deBeule2015}. But also for planetesimals
in a gaseous protoplanetary disk with gas pressures
reaching up to Martian values, dust ejection has the
capability of eroding their surfaces on orbits close to
the star \citep{Wurm2007x}. In addition, the ejected dust might complement
dust from collisions and be one component of the
dust observed in protoplanetary disks over their entire
lifetime \citep{deBeule2013}. Thermal creep might also drive gas accretion
in protoplanetary disks in combination with photophoresis.
As particles at the inner edge of disks are pushed
outward by the light, the gas moves inward toward
the star \citep{Kelling2013acc}.

\begin{figure*}[]
\epsscale{1.2}
\plotone{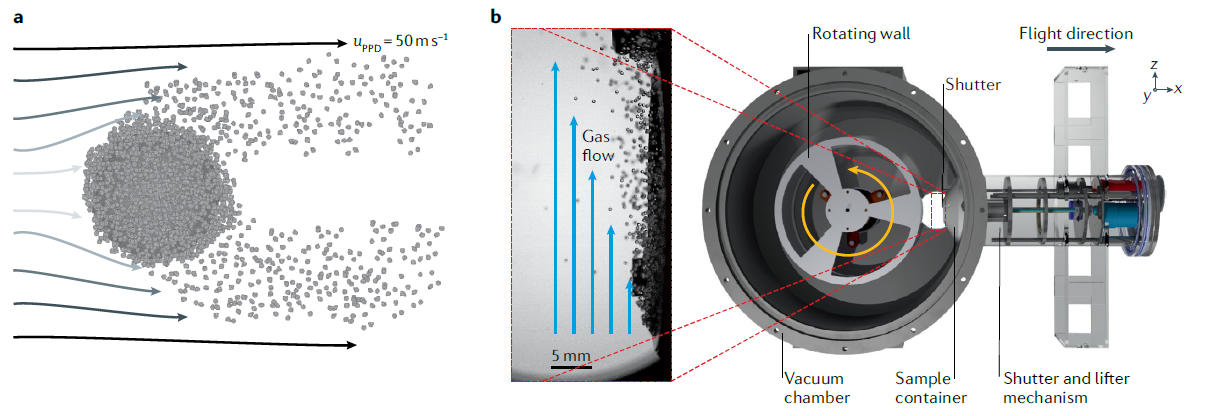}
\caption{\small{{Wind erosion and its experimental measurement.} \textbf{a}| Schematic of a pebble- pile planetesimal being eroded
by wind as it moves through the protoplanetary disk at velocity $u_{PPD}$. \textbf{b}| Schematic of a cutting- edge wind tunnel set up
to work in parabolic flights at low gravity and under low pressure, with a snapshot of eroded particles on the left side.
Part \textbf{a} adapted with permission from \citep{Demirci2019a}, OUP. Part \textbf{b} adapted with permission from \citep{Demirci2020}, OUP.}
\label{fig:5}}
\end{figure*}

\noindent
\textit{\textbf{Erosion by wind.}} Wind is a classical erosion mechanism
(and we note that we really do mean wind here,
not phenomena such as particle radiation in the form of
stellar winds). This subfield in planetesimal research is
currently gaining importance. In principle, it should be
on the topic list, because the current models of planetesimal
formation described above make these objects very
fragile pebble piles. In protoplanetary disks, a planetesimal
encounters head winds on the order of 50\,ms$^{-1}$ on
circular orbits, or stronger on eccentric orbits. Although
the pressure might be low, so is the planetesimal’s gravity,
and pebble piles in particular might be eroded
easily \citep{Demirci2019a,Rozner2020}.
Ground- based wind tunnel experiments at low pressure
with planetesimal erosion in mind found that planetesimals
in the inner region of the disk can be eroded
easily \citep{Paraskov2006}. These experiments still used compacted dust
samples instead of pebble piles, which were not the
favoured model of planetesimals at that time. However,
even compact dust planetesimals are prone to erosion,
especially if they move on eccentric orbits \citep{Schaffer2020}.
Planetesimals consisting of compact millimetre-size
dust aggregates (also called pebble piles) have only small
tensile strengths \citep{Skorov2012} when formed by hydrodynamic and
gravitational concentration and contraction from aggregates
resulting from the bouncing barrier. Measurements
on disintegrating dust aggregates with large constituents
in free fall are in agreement with this idea \citep{Musiolik2017}.
Such pebble piles cannot be studied easily on the
ground as gravity dominates over shear stress in holding
these pebbles down, and because gas drag at low
pressure is small. Therefore, a number of microgravity
experiments have been devised for parabolic flights.
A centrifuge on parabolic flights at zero g has been used
to produce g- levels from 0g to 1g. In addition, the centrifuge
was a vacuum chamber that could be evacuated
to millibar pressure, which hosted a wind tunnel \citep{Kruss2020,Musiolik2018,Demirci2019a}.
Further advances have been made by using a shear- flow
wind tunnel, following early work \citep{White1987} but using a vacuum
system and using particle samples with low shear
strength \citep{Demirci2020}. Now, erosion of pebble- pile planetesimals
can be studied at the pressures of protoplanetary
disks (Fig. \ref{fig:5}).
These experiments show that erosion of pebble pile
planetesimals is relatively easy and does not allow
these unconsolidated objects to exist in a region that
corresponds approximately to the inside of Mercury’s
orbit, depending on disk model and size of the object \citep{Demirci2020}.
The situation gets worse - that is, the instability region
becomes several AU in size - if the objects are on eccentric
orbits \citep{Demirci2020a}. So far, these results make the picture of
planetesimal formation inconsistent to some degree, as
planetesimals that would be eroded immediately would
not form in the first place. Therefore, this instability is
an important topic for the near future.
Another situation in which wind erosion is relevant,
which exists in a different phase of planet formation,
is pebble accretion by larger objects. This accretion is
currently a favoured model for the growth of larger
planetary bodies - and is strongly influenced by wind
erosion. As objects approach a protoplanet they are
accelerated. As they speed up, gas drag increases and
they can be eroded \citep{Demirci2020eropebble}. Small debris from this process will
couple to the disk’s gas and be carried away by the wind.
They therefore will not be accreted, which reduces the
efficiency of the process strongly.

\noindent
\textit{\textbf{Impacts.}} Impacts range from gentle to catastrophic. We
restrict our discussion to low- energy impacts of small
projectiles, which might not be a danger to a planetesimal
on a global scale but may still be an erosive process.
Such impacts are slightly different to the collision
processes discussed above in the sense that a granular,
non- cohesive nature of the constituents is expected
to be present and to prevail after a collision. Collision
experiments in this context therefore include impacts
into sand- like material or collisions with dust aggregates
but at moderate speeds, which is again the domain of
microgravity experiments \citep{Murdoch2017,Colwell2003,Colwell2008,Brisset2018,Whizin2017,Hestroffer2019,Garcia2015,Murdoch2013b,Kollmer2016,Tell2020}.
Such experiments have shown that impacts of small
projectiles into granular beds are highly dissipative.
Although typically many particles are ejected, a large
fraction of the ejecta does not reach the escape velocity
of the parent body. However, most of the ejecta will still
be transported away by gas drag \citep{Demirci2019a}. Impacting projectiles
will therefore increase the erosion rate severely.

\section{Late evolution}\noindent
The process of impacts, discussed above, continues until
a time when the primordial gas of the disk is mostly dispersed
after a few million years \citep{Haisch2001}. This is not the end
of planet formation, as the evolution of the remaining
objects does not stop here. In fact, it is only at this point
that the final planets are shaped. Ongoing collisions
of planetesimals and the accretion of pebbles leads to
the formation of a smaller group of large objects, called
planetary embryos. These might already have formed
while the gas was still around \citep{Lambrechts2019}. These embryos must
still collide with each other and collect remnant material
to grow further \citep{Watt2021,Zhu2019}.
Terrestrial planets have few time concerns here,
and the last large collisions of planetary- sized objects
might only occur after 100 million years \citep{Izidoro2015}. However,
the last step to form gas giants requires that planets of
a few Earth masses already grew while the gas of the
disk was still around, because only then is it possible
for the planet to accrete the gas and become a gas
giant \citep{Ndugu2021,Pollack1996,Boley2016}. Opposed to this core formation scenario
is the gravitational instability scenario, in which gas
giants are thought to form directly from a dense disk
by gravitational instability \citep{Boss1997,Boley2010}. However, core accretion
is the favoured mechanism for gas giants that are not
too far out, like Jupiter or Saturn. In any case, the presence
of large planets stirs the remaining reservoirs of
planetesimal-size objects substantially.
Catastrophic collisions between larger objects lead to
a broad size distribution of solids, typically following a
power law \citep{Krivov2021,MacGregor2016}. The current population of small bodies in
the Solar System (asteroids and comets) is treated as a remnant
of this era. Many space missions including OSIRISREx,
Hayabusa I/II and the Rosetta mission allow detailed
studies of these bodies. The view is limited to the presentday
state, so the evolution between their formation and
the present has to be conjectured. Here, microgravity
and laboratory experiments come into play again.
Asteroids often show a distinct size segregation on
their surface, with ponds filled with fine regolith \citep{Hughes2008,Kanamaru2019},
pebble- dominated regions \citep{Dellagiustina2019} and regions of bare rock \citep{Kanamaru2019}.
Different mechanisms are discussed as a reason for this
segregation, and are a focus of microgravity experiments
and experiments under reduced gravity levels.
Size segregation due to the ‘Brazil nut' effect is proposed
as one mechanism. According to this scenario, small
impacts induce vibrations within the parent body, which
then lead to a size sorting effect within the regolith
bed \citep{Schraepler2015,Fries2018,Guettler2013}.
Impacts of small particles and the resulting splash
events are also treated as a mechanism that leads to size
segregation \citep{Shinbrot2017}. Impacts into fine- grained regolith are
highly inelastic \citep{Sunday2016,Murdoch2017,Bogdan2020,Kollmer2016}, so that the impact ejecta remain
in the vicinity of the impact site despite the small gravitational
acceleration. Impacts onto rocky material are
more elastic, so that the impacting particles reach the
escape velocity or reach other regions of the asteroid,
where they are finally stopped.
Regolith particles on planetary surfaces are mobilized
not only by impacts but also by particle charging.
In this scenario, interaction between the solar wind
(particle radiation), UV radiation and the dust particles
leads to charging of the particles. If the majority
of the grains have charges with the same sign, the
repellent electrostatic forces can exceed cohesion
between the particles \citep{Sickafoose2002,Sternovsky2002,Colwell2007,Carroll2020}. This mechanism has been
proposed as the reason for dust ponds on the asteroid
433 Eros \citep{Hughes2008}.

\section{Outlook}\noindent
Laboratory experiments - on ground and under microgravity
- have been and remain mandatory to quantify
parameters from simple sticking velocities over fragmentation
limits to erosion rates. Such parameters are
widely used in numerical simulations on particle growth.
The connection from collisional growth to hydrodynamical
evolution remains a puzzling field, and experiments
are ongoing to quantify the maximum size of aggregates
that can be grown one way or another. One of the most
promising mechanisms is charged aggregation, and
experiments on microgravity platforms with longer
duration (such as sounding rockets) are next steps in
the pipeline. Additionally, the field of erosion is currently
experiencing some momentum. Although the process
itself is destructive, it might still be a constructive part
of planet formation. Here, experiments are required with
much lower gravity levels than provided by parabolic
flights and longer duration than provided by drop towers.
Overall, the processes discussed in this Review, from
collisions to erosion, often depend on the interaction of
many particles and a complicated mix of specific material
properties and conditions. Laboratory experiments
cannot stitch together the complex evolution from dust
to planets. However, they provide key input to numerical
simulations and information on critical crossings.

\section*{Acknowledgements}

A significant part of this work is supported by the German Space Administration (DLR) with funds provided by the Federal Ministry for Economic Affairs and Energy (BMWi) under grants 50WM1760, 50WM1762, 50WM2140, 50WM2142, 50WM2049. The authors acknowledge access to microgravity platforms in recent years by ESA. Part of the work is also funded by the German Research Foundation (DFG) under grants WU 321/16-1 and WU 321/18-1.

Nature Reviews Physics thanks Barbara Ercolano, Meiying Hou, Matthias Sperl and Ralf Stannarius for their contribution to the peer review of this work.

\newpage

\appendix

\section{Particle motion in protoplanetary disks} 
\label{ch:particlemotion}
\begin{figure*}[ht!]
\plotone{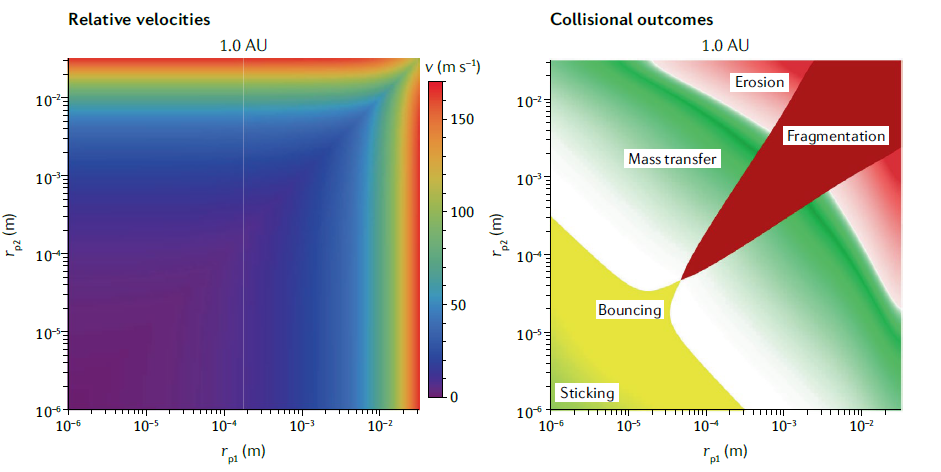}
\caption{\label{fig:bouncetofrag}\small{Figure adapted with permission from \citet{Husmann2016}, IOP.}}
\end{figure*}
\noindent
Particle motion in protoplanetary disks is strongly coupled to the gas motion, and thus gas drag on grains is substantial.
An important quantity is the gas–grain coupling time $\tau$, which quantifies the timescale on which a particle follows a change
in gas motion. It therefore appears in problems of defining absolute particle velocities and, more importantly, in relative
particle motion between two grains. For example, a grain above the midplane of a typical protoplanetary disk essentially
feels the gravitational force of the star. Whereas the radial component of the gravitational pull is balanced by the Kepler
rotation, its vertical acceleration $g_z$ leads to a settling of the grain with the terminal velocity $v_s = \tau g_z$. Two grains with
different coupling times $\tau_{1,2}$ then get a relative velocity and therefore a collision velocity of $v_c = |(\tau_1-\tau_2)| g_z$.
The coupling time is constant only if the drag force $F$ is proportional to the relative velocity between gas and particle, $v_{rel}$,
and it can be written as $F = (m/\tau)v_{rel}$ with particle mass $m$. For different drag regimes, $\tau$ can be calculated. For the well known
Stokes drag, $F_S = 6\pi\eta r v_{rel}$ with particle radius $r$ and gas viscosity $\eta$. The resulting coupling time is
\begin{equation}
    \tau_S=\frac{2}{9}\frac{\rho_P}{\eta}r^2
\end{equation}
with particle mass density $\rho_p$.
Stokes drag is relevant for flow that can be described by continuum hydrodynamics. In protoplanetary disks, this
description only holds for larger grains; the interaction of small grains is of a molecular nature. In the free molecular flow
regime, the coupling time is
\begin{equation}
    \tau=\frac{4}{3}\frac{\rho_P r}{\rho_g v_g}
\end{equation}
with gas density $\rho_g$ and thermal velocity of the gas molecules $v_g$. These relations are for isolated (test) particles that do
not react back on the gas. Otherwise, corrections are needed\citep{Schneider2019a,Schneider2019b}. We also note that these coupling times are based on compact, non-porous particles. For particles with fractal dimensions $d \leq 2, \tau$ does not depend on the size but remains constant.
Using typical disk models (see Appendix \ref{ch:protodisks}), $\tau$ can readily be calculated, and collision velocities can be quantified. The left part
of figure \ref{fig:bouncetofrag} shows relative velocities (in m\,s$^{-1}$) in a protoplanetary disk (model of LkCa15) at 1\,AU radial distance to the star,
assuming two particles of radius $r_{p1}$ and $r_{p2}$. The right part of the figure \ref{fig:bouncetofrag} shows collisional outcome regimes at the given sizes
and velocities \citep{Husmann2016}. Here, the colour code describes the different collision outcomes occurring in these regimes, ranging from
sticking, bouncing, mass transfer (from a smaller to a larger agglomerate) or erosion (of a larger agglomerate) to catastrophic
fragmentation (of both collision partners).
Besides the example of sedimentation given above, relative motion is also induced by Brownian motion (limited to dust),
turbulence and drift in radial and transverse directions. Radial drift is induced as the disk’s gas is supported by
a pressure gradient and rotates slower than the corrsponding Kepler velocity. For solid grains, which do not feel this
pressure gradient, this rotation is too slow. Therefore, if their coupling time is smaller than an orbit, they simply
experience a residual gravity toward the star. If the particles are so large that they only couple to the gas on timescales
much larger than an orbital timescale, this drift decreases in velocity again. The ratio between $\tau$ and orbital time is
defined as the Stokes number St$ = \tau/t_{orb}$.
Radial drift of solids follows a pressure gradient, which usually points inwards and is largest for particles with St $\sim 1$.
Depending on the location in the disk and the disk model, this corresponds to bodies of scale 0.1–1\,m. These drift inwards
at ~50\,m\,s$^{-1}$ (approximately 1\,AU per 100\,yr). This inward drift is a problem, because particles would end up in the star on
short timescales. However, if the pressure profile of the disk instead contains a maximum, the drift vanishes or can even be
reverted to an outward drift, and particles collect at the pressure maximum.

\section{Simulations} \label{ch:simulations}\noindent
Planet formation takes place in the midplane of protoplanetary disks, where it is
enshrouded by dust. Even at radio wavelengths, only small (centimetre-sized) particles
can be observed. Only bodies of planetary size appear in observations as exoplanets.
Detailed models and simulations are therefore needed to understand the process of
planet formation from dust to planets. There are several methods used, depending on
the specific problem to be solved. These simulations provide a major part of what we
know about planet formation, but as this Review is focused on experiments, we give
only a short summary here to provide context.

\subsection{Disk simulations}\noindent
The motion and evolution of a gaseous protoplanetary disk of several 100 AU in
diameter is essentially a very large hydrodynamic problem. Because the gas is partially
ionized, its motion also becomes sensitive to the stellar magnetic field \citep{Balbus1991}. Therefore,
(magneto)hydrodynamic codes are often used to study the principal motion of the
disk \citep{Kuiper2010,Stoll2016,Kuiper2018,Kley2012}. If global disk simulations are not sufficient to resolve the problems at hand
(for instance, details of streaming instabilities), simulations are often confined to small
sections of the disk, called shearing boxes, at a certain distance from the star \citep{Johansen2007,Chiang2010}. The
more the focus of a study shifts to embedded solids, the fewer details of the gas motion
can be simulated directly at the same time or on a larger scale. Disk models are then
mostly parameterized, and the motion and aggregation of solids and formation of
planetesimals within the disk can be predicted \citep{Zsom2010,Drazkowska2021,Windmark2012,Dullemond2018,Lenz2020,Drazkowska2017}.

\subsection{Collision simulations}\noindent
One feature common to all disk evolution models is that the outcome of interactions
between solid particles is not calculated directly, but included as input parameter.
Collision results therefore have to be provided by other means, in other words, by
zooming further into the disk. Experiments, as discussed in this Review, are one way to
gather empirical data on collisions or other processes. But not all parameter ranges are
accessible to experiments. Numerical simulations complement the picture and play an
important role in the understanding of planet formation.
There are essentially two types of collision simulations. Molecular dynamics
simulations treat dust agglomerates precisely, so each constituent particle is simulated
with its mechanical properties (density, surface energy, shape) \citep{Dominik1997,Wada2009,Wada2013,Okuzumi2012,Jankowski2012}. However, there
are size limits for these computations. Smoothed particle hydrodynamics has been
used for larger bodies. Strange as it sounds, a hydrodynamics code is used to simulate
solid bodies, which is described by a reservoir of elements carrying the macroscopic
mechanical properties (such as Young’s modulus or tensile strength) of the material.
This allows a dynamic description of collision processes, such as fragmentation,
compaction or bouncing.\\ 

\noindent
Even very large, porous agglomerates from metre to
planetesimal size can be simulated in this way \citep{Benz1988,Geretshauser2011,Meru2013}.

\section{Protoplanetary disks} 
\label{ch:protodisks}
\begin{figure*}[]
\plotone{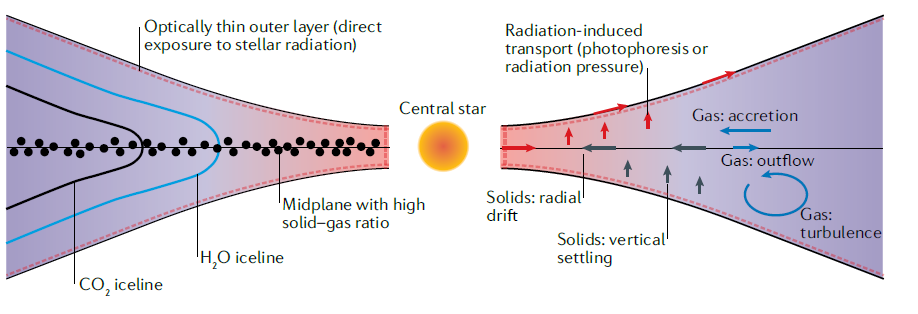}
\caption{\label{fig:diski}}
\end{figure*}
\noindent
Protoplanetary disks consist of essentially two phases, gas on the one hand and solids embedded within on the other
hand. Typically, the gas density and therefore gas pressure is higher closer to the star. In the simplest models, the radial
profile of the gas density in a disk is described by a simple power law, for example. The star in the centre continues to
grow by accreting material from the disk. During this accretion, potential energy of the infalling material is released as
heat. The absorption of stellar radiation adds to this. These heat sources result in specific vertical and radial temperature
profiles. In general, the temperature drops with distance from the star. At a characteristic radial distance and height
above the midplane, phase transitions of volatile materials occur - that is, sublimation and condensation of water,
$\rm CO_2$ or CO. These regions typically are referred to as snowlines or icelines. For major components such as water, this
phase transition strongly increases the density of solids on the cold side of the snowline. In addition, the pressure profile
can be modified locally, as the gas density is enhanced when the concentration of solids suddenly drops. The left part of
 figure \ref{fig:diski} gives an overview of the temperature profile, including some icelines.
To understand planetesimal formation as evolution of the solid component of protoplanetary disks, various dynamic
processes have to be considered. This dynamics is mainly driven by their interaction with the gas (Appendix \ref{ch:particlemotion}). The pressure
profile inflicts a radial drift on the solids, as the gas is supported against stellar gravity whereas the high- density solids are
not. In general, the drift of solids follows the local pressure gradient. While particles drift, they collide, possibly grow, and
change their composition. Thus the mechanical properties of solids change continuously during evolution. The right part
of figure \ref{fig:diski} gives some of the transport mechanisms. Additionally, stellar radiation and thermal radiation from the disk
lead to material transport by photophoresis or radiation pressure. \\

\bibliography{references}{}
\bibliographystyle{aasjournal}



\end{document}